# Dielectric permittivity tensor dynamics of in-plane hyperbolic van der Waals MoOCl$_2$ and emergent chiral photonic applications

A. V. Margaryan$^{1,+}$, M. L. Sargsyan$^{1,+}$, M. H. Hayrapetyan$^1$, D. A. Karakhanyan$^1$, K. S. Novoselov,$^{2,3}$
& D. A. Ghazaryan$^{1,*}$

$^1$Laboratory of Advanced Functional Materials, Yerevan State University, Yerevan, 0025, Republic of Armenia

$^2$Institute for Functional Intelligent Materials, National University of Singapore, Singapore, 117575, Republic of Singapore

$^3$Materials Science and Engineering, National University of Singapore, Singapore, 117575, Republic of Singapore

The correspondence should be addressed to: davitghazaryan@ysu.am

**Van der Waals (vdW) crystals offer unique opportunities for modern nanophotonic applications owing to their intrinsic anisotropic nature. While most of them exhibit uniaxial anisotropy arising from weak out-of-plane vdW interaction, some of their representative families also exhibit an in-plane biaxial anisotropy. Among the latter, outstand vdW oxochlorides with in-plane axes of a different physical character (metallic or dielectric). Here, we present an accurate dynamics of dielectric permittivity tensor components of vdW MoOCl$_2$ in the ultraviolet (UV) to visible (Vis) spectral region partly covering near-infrared (NIR). Addressing its enormously anisotropic optical constants, we focus on another hyperbolicity window of vdW MoOCl$_2$ emerging in the UV spectral region that may potentially unlock rich light-matter interaction effects. Furthermore, we propose an approach towards designing nanoscale handedness preserved Vis light circular polarizers based on twisted helical vdW MoOCl$_2$ heterostructures. Our findings display that vdW MoOCl$_2$ provides a highly promising platform not only for hyperbolic, but also for chiral nanophotonic applications.**

**Introduction**

Van der Waals crystals form a rapidly expanding class of layered solids nowadays comprising a vast variety of representative families[1–5]. Since the isolation of graphene[6,7] and the following emergence of transition-metal dichalcogenide monolayers[8–10], a plethora of an exciting and tuneable properties were uncovered in these layered crystals, such as thickness-dependent bandgaps[11,12], strong excitonic effects[13,14], polarization-selective responses[15], as well as more sophisticated phenomena including natural hyperbolic polaritons[16,17], topological electronic phases[18,19], intrinsic 2D magnetic states[20,21], *etc.*. Furthermore, the ability to cleave an ultrathin layer from a vast variety of vdW crystals of different physical character and assemble it into an artificial vdW heterostructure[22–24] with atomically clean interfaces allowed twist-induced moiré excitonic[25] and polaritonic[26] states to manifest opening novel pathways for designing of nanoscale optoelectronic and photonic functional devices[27,28].

Transition-metal oxyhalides with the general composition of *MOX*$_2$ (*M* = transition metal, *X* = halogen) recently emerged as a quite promising family of vdW crystals exhibiting strong biaxial anisotropic properties especially pronounced for the in-plane components[29]. Its several representatives[30,31,32], such as vdW NbOCl$_2$, NbOI$_2$, VOCl$_2$, VOBr$_2$, and MoOCl$_2$ host pronounced structural, and hence, also an optical anisotropy, whereas for vdW RuOCl$_2$ and OsOCl$_2$, the current understanding is yet limited mainly to a computational prediction[33, 34]. A subgroup of these vdW oxochlorides comprising $NbOCl_2$, $MoOCl_2$, $RuOCl_2$, and $OsOCl_2$ shares a characteristic structural motif of metal-oxygen chains coordinated by halide atoms. Such mixed metal-oxygen directional bonding architecture produces pronounced in-plane anisotropy making representative vdW low-symmetry crystals an attractive platform for the investigation of an unconventional dispersion rooted nanophotonic phenomena. In the case of vdW $NbOCl_2$ and $MoOCl_2$, previous studies have shown an enormous in-plane optical anisotropy along with a hyperbolicity of the spectrum that allows high-$k$ sub-diffraction polariton propagation effects in UV ($NbOCl_2$) and NIR spectral regions ($MoOCl_2$)[35–39].

In this manuscript, we systematically approach the anisotropic optical properties of hyperbolic vdW $MoOCl_2$ in the 250-1000 nm spectral region. Our studies encompass a symmetry-based resolution of polarization sensitive properties of Raman active vibrational modes along with an accurate in-plane anisotropic optical constants. Notably, even a slightest inexactitude in optical constants is crucial for a vast variety of emergent nanophotonic applications[40]. Our work sheds new light not only on an accurate dynamics of dielectric tensor components and the precise hyperbolic transition wavelengths extracted through a direct study but also displays an emergence of another hyperbolic window of vdW $MoOCl_2$ in UV spectral region. Besides, leveraging the enormous in-plane anisotropy of vdW $MoOCl_2$, we demonstrate that its twisted bilayers exhibit strong chiral optical properties in the Vis spectral region paving the way to all-vdW chiral photonic applications.

**Results**

**Angle-resolved vibrational characterization of vdW $MoOCl_2$**

The crystal structure (see Fig. 1a) of vdW $MoOCl_2$ belongs to the monoclinic $C_2/m$ space group[35] suggesting a presence of a strong structural anisotropy[36,37]. It is a biaxial crystal characterized by the metallic Mo-O chains and dielectric Mo-$Cl_2$ linkages that result in the emergence of an enormous in-plane anisotropy. Therefore, the cleaved samples typically display an elongated morphology with their long and short edges aligned to crystallographic $x$ and $y$ axes. Here, approaching as in[41], one may directly study the in-plane contribution of this anisotropic response with regard to the vibrational modes by the employment of an angle-resolved Raman spectroscopy (see Fig. 1b and 1c) extending polarization sensitive response of vdW $MoOCl_2$ above (see Fig. 1d and 1e) the nominal in-plane $x$ and $y$ axes studies[42]. Capturing the complete polarization response over a full sample rotation range (0-360°) in both co-polarized ($e_i \parallel e_i$) and cross-polarized ($e_i \perp e_s$) (see Supplementary Note 1) configurations allows to see a direct correlation between the Raman intensity, sample orientation, the physical character of the crystallographic axes and extract the necessary information for an unambiguous determination of the symmetry of each Raman mode. Our studies reveal six Raman modes in 100-500 cm$^{-1}$ spectral region, where one belongs to $B_g$ (124.5cm$^{-1}$) anti-symmetric group (see Fig. 1c) and the other five to the symmetric $A_g$ (178.5 cm$^{-1}$, 293 cm$^{-1}$, 309 cm$^{-1}$, 350.5 cm$^{-1}$ and 430 cm$^{-1}$) group. The vibrational modes of vdW $MoOCl_2$ can be analysed by a group theory based on $C_2^h$ point symmetry within the framework of Placzek's approximation[43], Fig. 1f. Here, Raman-active modes that correspond to the irreducible representations of $A_g$ and $B_g$ modes can be described by symmetry-restricted $\alpha$ tensors as:

$$\alpha(A_g) = \begin{pmatrix} a & 0 & 0 \\ 0 & b & d \\ 0 & d & c \end{pmatrix}, \qquad \alpha(B_g) = \begin{pmatrix} 0 & e & f \\ e & 0 & 0 \\ f & 0 & 0 \end{pmatrix}, \tag{1}$$

where $a$, $b$, $c$ $d$, $e$ and $f$ are the symmetry elements. Note that $x$ is taken here as the unique axis and $yz$ as the mirror plane (hence, the space group is mirrored to $B_2/m$). The scattering amplitude, and therefore, the intensity of Raman response can be expressed as $J \propto |e_s^T \alpha(j) e_i|^2$, where $e_i$ and $e_s$ are polarization unit vectors of incident and scattered lights. In backscattering regime $e_i(\vartheta) = (\cos(\vartheta),\sin(\vartheta), 0)$ and $e_s^{\parallel}(\vartheta)= e_i(\vartheta)$, where $\vartheta$ is the rotation angle between the incident polarization and one of the crystallographic axes (dielectric $y$-axs in our case).

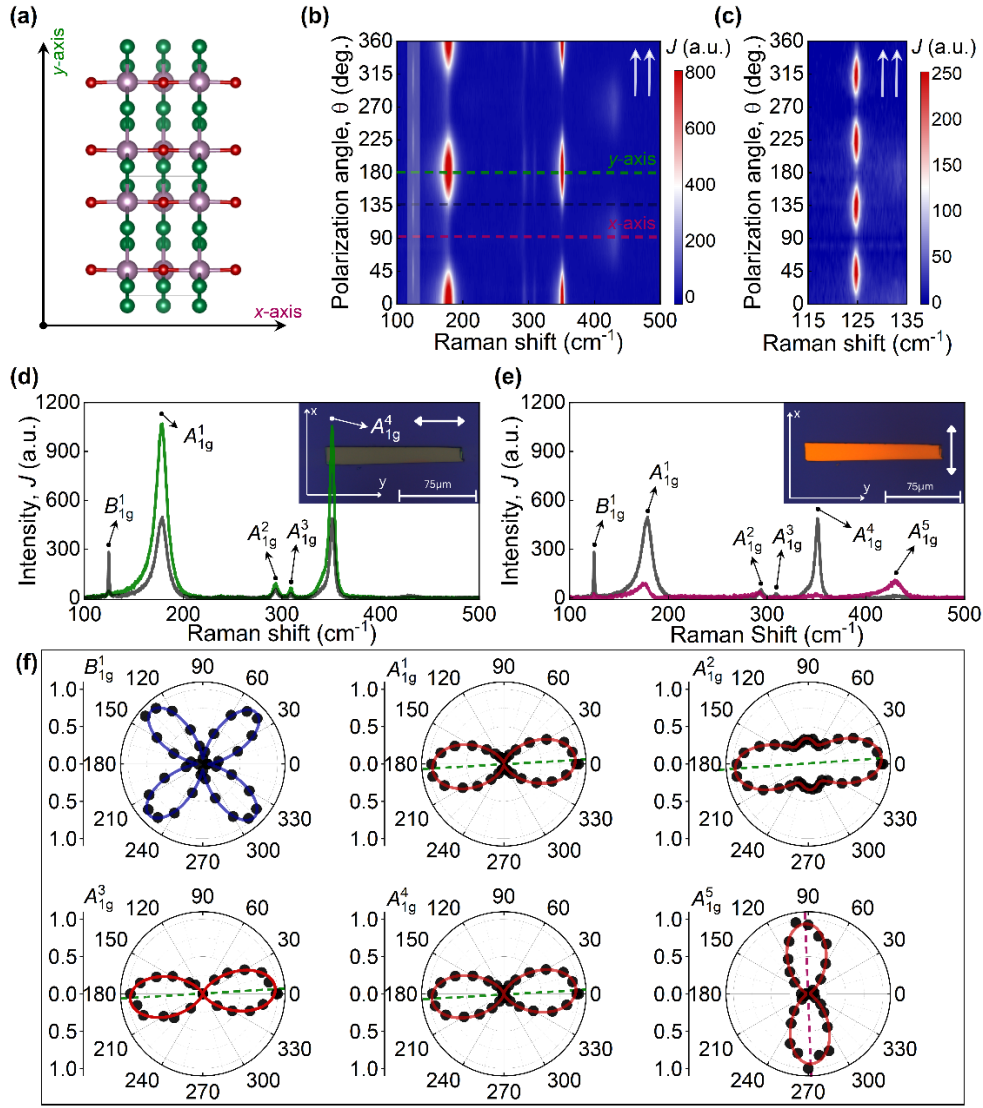

**Fig. 1 | Polarization dependent Raman mapping of metal-to-insulator in-plane axes of vdW MoOCl$_2$. a** Crystal structure of monoclinic vdW MoOCl$_2$ (space group $C_2/m$) emphasizing the crystallographic $x$-axes (Mo-O chain direction) and $y$ (Mo-Cl$_2$ linkage direction). **b,** Angle-resolved Raman intensity map acquired in the co-polarized ($e_i \parallel e_i$) configuration within 100-500 cm$^{-1}$ spectral region. The dashed green and red lines emphasize the crystallographic $y$ and $x$ axes positions. **c,** Zoomed-in part of the map (see shaded white rectangle) presented in (b) showing the angular evolution of $B_g$ mode at 115-135 cm$^{-1}$ spectral region. **d,** and **e,** Raman spectra plots acquired along in-plane dielectric $y$-axis (green curve, (d)) and metallic $x$-axis (red curve, (e)). The black curve in (d, e) demonstrates the Raman response acquired at 45° alignment. The insets in (d, e) display the iridescent optical micrographs of the measured vdW MoOCl$_2$ sample taken with LED light when an incident polarization is aligned with corresponding in-plane crystallographic $x$ and $y$ axes. **f,** Polar diagrams of the representative $A_g$ and $B_g$ Raman modes shown as a function of polarization angle $\vartheta$.

The substitution of these polarization unit vectors into (1) yields the following angular dependencies of $J_\parallel^{A_g} \propto a^2 [(\sin^2(\vartheta)+(b/a)\cos(\varphi_{ba})\cos^2(\vartheta))^2+((b/a)\sin(\varphi_{ba})\cos^2(\vartheta))^2]$ for $A_g$ modes, and $J_\parallel^{B_g} \propto e^2\sin^2(2\vartheta)$ for $B_g$ mode, where $\varphi_{ba}$ is the phase difference between $a$ and $b$ symmetry elements. As a result, we see (see Fig. 1f) that the angular variation of $A_g$ modes exhibits distinct polarization dependencies based on relative magnitudes of $a$ and $b$ symmetry elements. Here, for $A_1$, $A_3$ and $A_4$ modes $a \gg b$ condition results in a two-lobed intensity pattern with a maxima when the incident polarization is aligned along the dielectric $y$-axis. In contrast, $A_5$ mode, where $b \gg a$, exhibits a two-lobed pattern rotated by 90° reaching maximum intensity along the metallic $x$-axis. Such behaviour allows $A_g$ modes to be used as reliable indicators of the crystallographic axes in vdW $MoOCl_2$ samples. $B_g$ mode, on the other hand, displays a symmetric four-lobed angular dependence rotated by 45° with respect to $A_g$ modes, which is consistent with its off-diagonal tensor character confirming the orthogonal polarization selection rules (see Supplementary Note 1 for the measurements in cross-polarization configuration).

**Ultraviolet and visible to near-infrared hyperbolic spectral regions of vdW $MoOCl_2$**

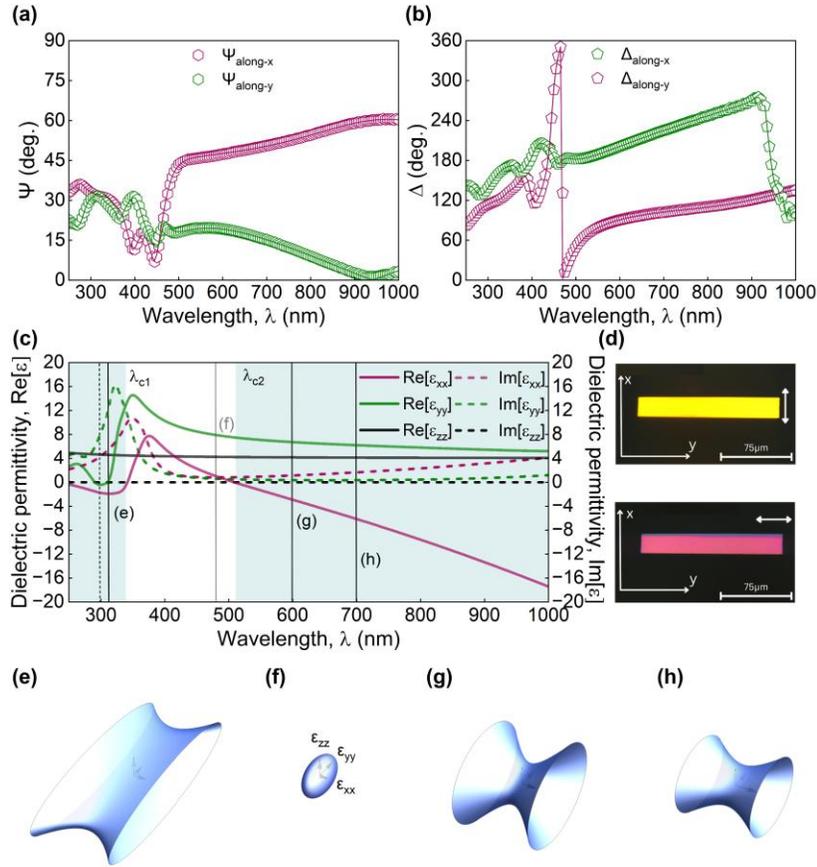

**Fig. 2 | Ultraviolet, visible to near-infrared hyperbolic regions and axis-dependent dielectric permittivity tensor components of vdW $MoOCl_2$. a,** Ellipsometric parameter $\Psi$ acquired at 55° angle of incidence (AOI) along the metallic $x$ and dielectric $y$ in-plane axes. **b,** Same as (a), but for ellipsometric parameter $\Delta$. **c,** The real and imaginary (Re[$\varepsilon$] and Im[$\varepsilon$]) parts of the dielectric permittivity tensor components along $x$, $y$, and $z$ crystallographic axes. The shaded regions below $\lambda_{c1}$ = 341 nm and above $\lambda_{c2}$ = 509 nm indicate spectral regions, where $\varepsilon_{xx} < 0$, that correspond to the first and second hyperbolic windows, while the intermediate unshaded region represents the standard dielectric window ($\varepsilon_{xx} > 0$). The dashed black line emphasizes the narrow region with a Type-$II$ hyperbolic window. **d,** The iridescent optical micrographs of a representative vdW $MoOCl_2$ sample taken with LED light when an incident polarization is aligned with the corresponding in-plane crystallographic $x$ and $y$ axes. **e, g** and **h,** The calculated isofrequency surfaces in $k$-space at representative

wavelengths of 320 nm, 600 nm and 700 nm corresponding to the both hyperbolic windows ($\varepsilon_{xx}$ < 0, $\varepsilon_{yy}$, $\varepsilon_{zz}$ > 0), while **f,** represents the standard ellipsoid surface at 480 nm corresponding to the dielectric window.

After the determination of a metallic *x* and dielectric *y* crystallographic axes of a representative vdW MoOCl$_2$ sample, we performed spectroscopic micro-ellipsometry (see Methods) studies orienting one in-plane axis of it in such a way to enable reliable extraction of the ellipsometric parameters *Ψ* and *Δ* along the crystallographic *x* and *y* axes (see Fig. 2a and 2b). The latter exhibits strong direction dependent variations clearly revealing not only the enormous in-plane optical anisotropy of vdW MoOCl$_2$, but also the metallic and dielectric responses of the crystallographic axes. Here, in Vis and NIR spectral regions, *Ψ* parameter responses are significantly higher for the metallic *x*-axis if compared to dielectric *y*-axis, whereas *Δ* parameter exhibits an opposite behaviour displaying a plateau for a metallic *x*-axis in case of relatively interference free thicknesses of vdW MoOCl$_2$ samples (see Methods, Supplementary Notes 2 and 3 for the extended set of data and further employed Muller-Matrix (MM) micro-ellipsometry along with the corresponding data). The retrieved complex permittivity tensor components along the *x*, *y*, and *z* axes (see Fig. 2c) reveal two distinct hyperbolic regions, where Re[$\varepsilon_{xx}$] < 0 below $\lambda_{c1}$ = 341 nm and above $\lambda_{c2}$ = 509 nm. Those are separated by a narrow dielectric region, where all principal dielectric permittivity components are positive. The iridescent micrographs shown in Fig. 2d for a representative vdW MoOCl$_2$ sample are a direct consequence of an enormous in-plane optical anisotropy in the second hyperbolic window.

The set of the allowed wavevectors that can propagate inside an anisotropic media is often envisaged by the isofrequency surfaces[44]. When all the principal components of dielectric permittivity tensor are positive, this surface forms a conventional index ellipsoid. Whereas when one or two tensor components get negative, it transforms into a hyperboloid, which can be classified as Type-*I* or Type-*II*. Fig. 2e-h illustrate the evolution of these isofrequency surfaces for an in-plane anisotropic vdW MoOCl$_2$ as the wavelength varies. Here, in the UV spectral region (see Fig. 2e), where Re[$\varepsilon_{xx}$] < 0 and Re[$\varepsilon_{yy}$], Re[$\varepsilon_{zz}$] > 0, vdW MoOCl$_2$ exhibits a Type-*I* hyperbolic dispersion along with a narrow region of wavelengths with Type-*II* dispersion, where also Re[$\varepsilon_{yy}$] < 0 at the vicinity of $\lambda_{c\text{-Type-II}}$ = 299 nm. The former hyperbolic isofrequency surface plotted in *k*-space (where $k = qc/\omega$) is shown in Fig. 2e. Between 341 nm and 509 nm, all dielectric tensor components become positive and the surface transitions into a conventional ellipsoid, Fig. 2f. At the longer wavelengths above 509 nm, Re[$\varepsilon_{xx}$] again becomes negative, restoring Type-*I* hyperbolic dispersion. The resulting isofrequency surfaces at 600 nm and 700 nm (see Figs. 2g and 2h) clearly display the re-emergence of the hyperbolicity in the Vis to NIR spectral region.

Extracting the refractive indices ($n_i$) and extinction coefficients ($k_i$) for all crystallographic axes from dielectric permittivity data (Fig. 3a), we observe that the extinction coefficient along the metallic *x*-axis reaches values up to 4.2 in the NIR spectral region (at 1000 nm), while it remains close to zero along the out-of-plane *z*-axis, and almost infinitesimally low (0.26) along the dielectric *y*-axis. In contrast, the refractive index along the dielectric *y*-axis increases up to 3.2 towards the inception of Vis region (at 400 nm), whereas for the metallic *x*-axis it remains below unity. Along *z*-axis, the refractive index is 2.1 exhibiting a weak dispersion across the measured spectral region (see tabulated dielectric permittivity tensor components and optical constants in Supplementary Note 4).

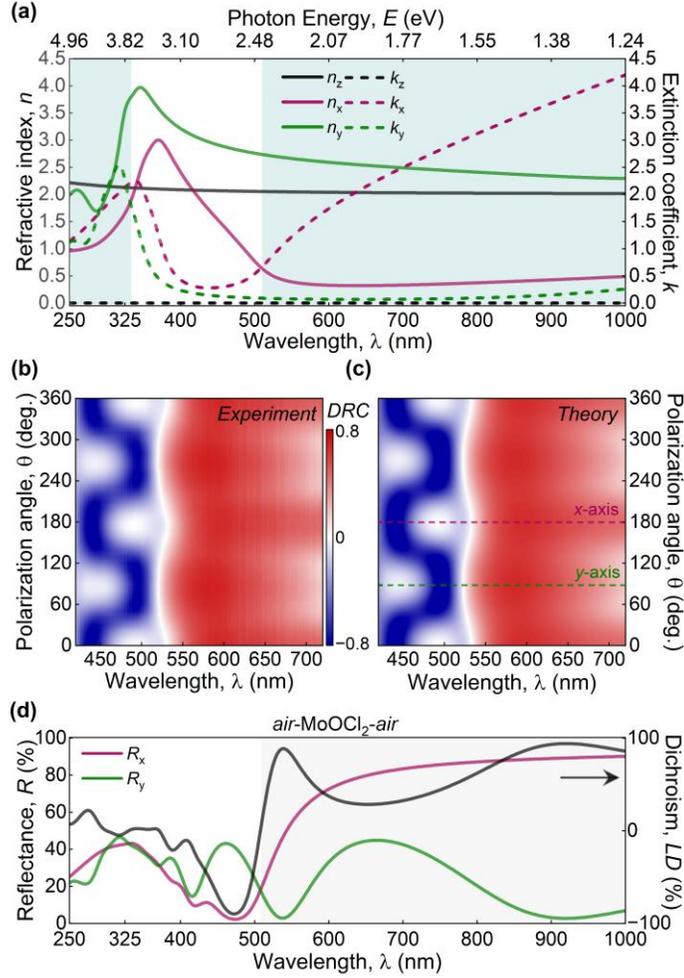

**Fig. 3 | Enormous in-plane optical anisotropy between metal-to-insulator in-plane axes of vdW MoOCl₂. a,** The optical constants ($n_i$, $k_i$) of vdW MoOCl₂ acquired along *x*, *y*, and *z* crystallographic axes showing an enormous optical anisotropy, though a lossy one along the metallic *x*-axis. The shaded areas correspond to two Type-*I* hyperbolic regions. **b,** Experimental angle-resolved differential reflectance contrast (DRC) map of a representative MoOCl₂ sample cleaved onto a Si/SiO₂ substrate. **c,** Theoretical angle-resolved DRC map calculated using the acquired optical constants shown in (a). The dashed green and red lines emphasize the crystallographic *x* and *y* axes. **d,** Simulated reflectance spectra of a 200 nm vdW MoOCl₂ sample for the cases of an incident polarization subsequently aligned with *x* and *y* axes. The shaded region above 509 nm indicates the second Vis to NIR hyperbolic spectral region of vdW MoOCl₂, where the polarization dependent contrast becomes extremely pronounced. The black curve depicts the linear dichroism (*LD*) in reflectance responses between the metallic *x* and insulating *y* axes.

To further validate our findings, we performed micro-reflectance spectroscopy of vdW MoOCl₂ samples (see Methods for more details). Here, unlike the latter operating at an oblique incidence, normal-incidence reflection studies provide complementary information about the intrinsic optical response of a novel material. The angle-resolved differential reflectance contrast (DRC) map measured for a representative vdW MoOCl₂ sample on a Si/SiO₂ substrate is shown in Fig. 3b. Using the retrieved optical constants extracted from micro-ellipsometry, we also calculated the DRC map (see Fig. 3c). Here, the difference between the experimental and simulated values does not exceed 3 % confirming the high accuracy of the extracted data (see the complete set of data in Supplementary Note 5). Furthermore, using these optical constants, the reflectance *R* of a vdW MoOCl₂ was calculated for an *air*-MoOCl₂-*air* system to illustrate its intrinsic reflectivity properties. As displayed in Fig. 3d, for a 200 nm thick vdW MoOCl₂ layer, the simulations reveal an enormous optical anisotropy with a pronounced reflectance polarization contrast between the metallic *x* and dielectric *y* axes in the second hyperbolic region. To provide its quantitative measure, we calculated the linear dichroism defined as $LD = (R_x - R_y)/(R_x + R_y)$ and observed that it is remarkably large reaching 88 % in the vicinity of second hyperbolic transition wavelength of $\lambda_{c2}$ = 509 nm and further rising to 95 % in NIR spectral region (see the complete set of evaluated data for other thicknesses in Supplementary Note 6).

**Emergent chirality established on hyperbolic vdW MoOCl$_2$ bilayers**

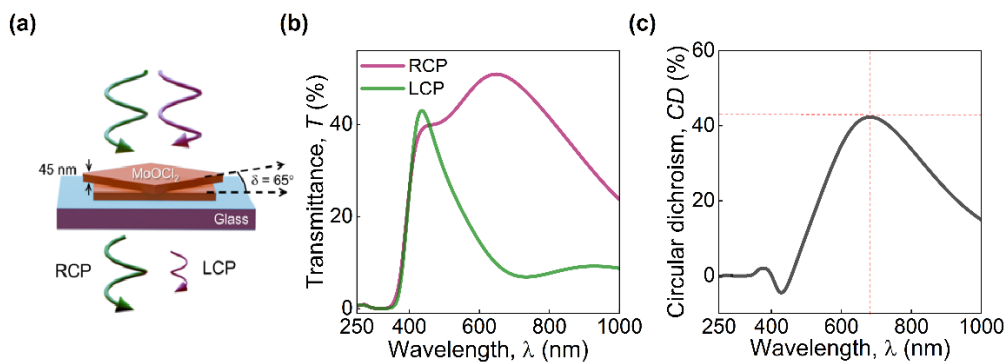

**Fig. 4| Towards all-vdW handedness preserved circular polarizers with helical heterostructures based on in-plane hyperbolic MoOCl$_2$. a,** Schematic design of a bilayer vdW heterostructure positioned on top of a conventional glass substrate. RCP and LCP stand for the illuminated (from the top) right-circular and left-circular polarized lights. **b,** Calculated transmittance of RCP and LCP lights traversed through vdW heterostructure depicted in (a). **c,** Calculated circular dichroism (*CD*) for RCP and LCP lights shown in (b).

The enormous in-plane optical anisotropy of vdW MoOCl$_2$ in Vis to NIR spectral region makes it an attractive candidate for the polarization-sensitive and chiral nanophotonic applications. Building on this anisotropy, we explore the chiral optical properties of the twisted vdW MoOCl$_2$ heterostructures, where constituent layers are twisted and crystallographically aligned to a certain degree forming a helical stack (see Supplementary Note 7). Remarkably, a strong chiral response emerges already for a bilayer vdW heterostructure (see Fig. 4a) enabling the design of ultrathin chiral optical components. Here, using the extracted anisotropic dielectric constants, we compute the circularly polarized transmission through a 65° twisted vdW MoOCl$_2$ heterostructure, where the constituent layers are 45 nm thick (see Fig. 4a). Fig. 4b demonstrates the calculated transmission spectra for the right-circularly (RCP) and left-circularly (LCP) polarized incident lights traversed through such a bilayer vdW heterostructure. Next, we calculate[45] the circular dichroism as $CD = T_{RCP} - T_{LCP}$, where $T_{RCP}$ and $T_{LCP}$ are the circular basis transmission elements (see Methods and Supplementary Note 7 for details). The resulting *CD* spectrum is displayed in Fig. 4c, which is showing an emergence of a strong twist-induced optical chirality reaching values as high as 42 % peaking at 682 nm red-light wavelengths. This pronounced selectivity of polarization may enable the designing of nanoscale handedness preserved circular polarizers potentially offering valve-like control over the RCP and LCP lights when the twist angle's helicity is controlled *in-situ*[46,47], which is quite feasible for a bilayer stack.

**Discussion**

When comparing vdW hyperbolic crystals, it is important to bear in mind that the negative dielectric permittivity may originate from various physical mechanisms, such as phonon-polariton resonances[48,49], plasmon-polariton (Drude-type) anisotropic free-carrier response[34,35] and anisotropic interband transitions[31]. In the phonon-polaritonic systems, the hyperbolicity emerges only within the mid-infrared Reststrahlen bands, where optical phonon resonances drive the sign reversal of the real part of the relevant permittivity component. In contrast, plasmon-polaritonic hyperbolicity emerges from anisotropic plasma frequencies along various crystallographic axes leading to an appearance of the negative real part of a relevant dielectric permittivity tensor component and often appears over a broad spectral range. Another route to hyperbolicity occurs when strongly directional interband transitions generate a resonant Lorentz contribution that drives one or more imaginary parts of dielectric permittivity tensor components to an enhancement, and hence, to a sign reversal of the real part of the corresponding component. Furthermore, a negative dielectric permittivity, and hence, hyperbolicity may occur in vdW crystals at either in-plane or out-of-plane directions depending on the symmetry of the dielectric tensor and irrespective of the formation mechanism. In the case of the phonon-polaritonic mechanism, among many others[16,17], one may separate hBN[48], which exhibits out-of-plane hyperbolicity, whereas $\alpha$-$MoO_3$ and $\alpha$-$V_2O_5$ host an in-plane hyperbolic phonon-polaritons all restricted to the mid-infrared spectral region[50,51]. Beyond the phononic mechanism, vdW crystals also host in-plane plasmonic hyperbolicity, for instance, vdW $WTe_2$ supports the plasmon-polaritons in the far-infrared spectral region[52] along with vdW $MoOCl_2$ that hosts them in Vis to NIR spectral region. A related plasmon-polaritonic hyperbolicity has also been predicted for another member of vdW family: $RuOCl_2$, within the UV spectral region[34]. For another instance, vdW $NbOCl_2$ displays hyperbolicity arising from the interband transitions generally occurring in the near blue to UV spectral region[31]. In the case of our vdW $MoOCl_2$, hyperbolic windows originate from two different mechanisms. In the UV region, the hyperbolicity is interband transition driven – a strong and highly anisotropic interband absorption peak along $x$-axis, Im[$\varepsilon_{xx}$], produces a negative real permittivity Re[$\varepsilon_{xx}$] via Kramers-Kronig relation. A similar, but much weaker resonance also causes Re[$\varepsilon_{yy}$] to cross slightly below zero over a very narrow spectral interval (see Fig. 2(c)) resulting in a short window of Type-*II* hyperbolic region. On the other hand, Vis to NIR hyperbolic window originates from an anisotropic plasmon-polaritonic free-carrier response that can be described by an anisotropic Drude term along the conductive Mo-O chains ($x$-axis).

The optical chirality may emerge in vdW heterostructures in accordance with a variety of mechanisms. Among the others, the configurational chirality is based on a certain geometry of the stacked layers rather than the intrinsic chirality of an individual constituent layer. In graphene and/or hBN based vdW heterostructures, the emergence of a chirality is of a microscopic origin as it is based on a helical symmetry of the formed moiré superlattices exhibiting small twist angles[53,54]. Twisted transition metal dichalcogenide bilayers as, for another instance, introduce the case of a different microscopic mechanism based on the valley-selection rule[55,56,57]. Configurational chiral optical effects also arise in vdW heterostructures when the adjacent optically thick layers are helically twisted with respect to each other. This mechanism is of a macroscopic origin and is based on an in-plane anisotropic properties of constituent vdW layers as it was shown for vdW $As_2S_3$[58]. The latter exhibits no absorption losses though is limited to a 10 % *CD* in the Vis spectral region according to the proposed model for a trilayer stack of 480 nm total thickness. Due to an enormous in-plane anisotropy, a 90 nm thick twisted bilayer vdW $MoOCl_2$ heterostructure already exhibits a strong chiral optical response displaying *CD* reaching up to 42 % for the transmitted signals. The calculated absorption of the handedness allowed (blocked) circular polarized light traversed through our bilayer stack is 21.2 % (29.5 %) at the vicinity of *CD*'s maximum wavelength indicating that it delivers strong chiral optical effects while sacrificing a portion of the signal to the absorption losses, which also exhibit *CD* (see Supplementary Note 7 for details).

**Methods**

**Sample preparation.** Bulk MoOCl$_2$ crystals were purchased from 2D Semiconductors as a custom-synthesized crystalline material and micro-mechanically cleaved into vdW samples on top of standard Si, Si/SiO$_2$ and quartz substrates at room temperature using commercial scotch tapes from Nitto Denko Corporation. Prior to the micro-mechanical cleavage, the substrates were sequentially cleaned in acetone, isopropanol and deionized water, which was followed by an air plasma treatment in order to remove residual surface contaminants and improve the adhesion.

**Angle-resolved Raman spectroscopy.** Raman spectra from vdW MoOCl$_2$ samples were acquired using a Horiba LabRAM HR Evolution confocal microscope equipped with a 100X objective lens (*N.A.* = 0.90) and an 1800 lines/mm (450-850 nm) diffraction grating. The measurements were carried out using a 633 nm excitation laser and SIN-EM FIUV type detector operating at -75 °C temperature. The initial position of vdW MoOCl$_2$ samples were aligned with one of the crystallographic axes as shown in the insets of Fig. 1d and 1e. The polarization angle was varied continuously from 0° to 360° with a step of 10°. The acquisition time was set to 10 s per point with 4 accumulations. The laser power on vdW MoOCl$_2$ samples were maintained at 34 mW using a 50 % neutral density filter to minimize a possible photoinduced damage. Spectral processing involved baseline subtraction followed by the fitting of Raman peaks with Gaussian functions. The extracted intensity values were then used to construct the angular dependent polar diagrams for each of the modes. For the fitting of the latter, a slight misalignment between the crystallographic axes and the laser polarization was accounted by an introduction of a small rotation offset of $\vartheta_0$. In Addition, a potential mismatch between the laser polarization and the analyzer orientation was included through an additional fitting parameter $r^2$ in the intensity functions.

**Spectroscopic micro-ellipsometry.** Spectroscopic micro-ellipsometry measurements were carried out at room temperature at 250-1000 nm spectral region using a Park Systems Accurion EP4 rotating-compensator imaging ellipsometer. All optical modeling and fitting procedures were performed using the EP4Model software package. Multiple vdW MoOCl$_2$ samples of varying thicknesses across various substrates (Si, Si/SiO$_2$ and quartz) were measured and analysed to ensure the reliability and enhance the accuracy of the obtained data. For each sample, the crystallographic *x* and *y* axes were first aligned with respect to the plane of incidence to enable an initial characterization using standard micro-ellipsometry. In this configuration, the ellipsometric parameters *Ψ* and *Δ* were collected at AOIs of 60°, 65°, and 70°. To verify the correct alignment, single-point MM micro-ellipsometry was performed confirming the negligible values of off-diagonal elements. An initial anisotropic model was then constructed and fitted using two in-plane and one out-of-plane components. The out-of-plane dielectric permittivity was described using a Cauchy dispersion model, $n(\lambda) = A + B\lambda^2 + C\lambda^4$. The in-plane response along the metallic *x*-axis was parameterized using a Drude-Lorentz model comprising an anisotropic Drude term and Lorentz oscillators while the dielectric *y*-axis was modeled using Lorentz oscillators only. An extensive MM micro-ellipsometry was next performed on an arbitrarily oriented vdW MoOCl$_2$ samples at 60°, 65°, and 70° AOIs across the 400-900 nm spectral region to extract the dielectric tensor components in full. The initial parameter set was then refined through the analysis of multiple samples and by minimizing the mean-squared error (MSE). An additional single-point MM measurements at a 532 nm incident wavelength were conducted at 65° AoI while rotating the sample in 0° to 180° range with a step of 10°. This measurement sequence allowed to reduce the parameter correlations and enable more accurate probing of the out-of-plane dielectric contribution. The optical constants of vdW MoOCl$_2$ were eventually obtained based on minimization of MSE of an optimized anisotropic model considering the examination of the parameter correlation matrix and accounting for the depolarization effects (see Supplementary Note 3 for further details).

**Micro-reflectance spectroscopy.** Micro-reflectance spectra were acquired using a Leica DM6M upright microscope equipped with a LED light source as well as a polarizer and an analyzer furnished for the incident light. The angle-resolved micro-reflectance spectra were recorded in 420-720 nm spectral region for a variety of vdW $MoOCl_2$ samples. Notably, the latter were aligned by their crystallographic *x*-axis to the parallel polarized spectroscopy configuration prior to measurements. Thorlabs CCS100/M spectrometer coupled to a 105 μm core diameter optical fiber was used for the spectral acquisition in combination with a 20X objective lens. This framework enabled spectral studies from a region of 10 μm in the lateral size[59]. The polarization dependent reflectance spectra were calculated using a 4 × 4 transfer matrix method accounting for the peculiar anisotropic properties of vdW $MoOCl_2$ and by defining propagation and boundary conditions for each interface (see Supplementary Note 5 for more details).

**Chirality calculations.** Chiral optical properties of twisted vdW $MoOCl_2$ bilayers were calculated for *air*-$MoOCl_2$-$MoOCl_2$-*glass*-*air* system using the generalized transfer-matrix method (GTM)[60]. These vdW heterostructures were modeled as anisotropic layers using the dielectric tensors extracted from micro-ellipsometry. Circularly polarized incident lights were constructed in the Jones basis, where propagation through vdW heterostructures were evaluated using the standard Maxwell boundary conditions for the corresponding propagation and interface matrices. Here, we calculated the transmitted lights for RCP and LCP signals and used those to evaluate corresponding *CD* (see Supplementary Note 7 for more details).